\documentclass[prd,singlecolumn,superscriptaddress,showpacks]{revtex4}
\usepackage[utf8]{inputenc}
\usepackage{empheq}
\usepackage{amsmath}
\usepackage{amssymb}
\usepackage{stmaryrd}
\usepackage{dsfont}
\usepackage{graphicx}
\usepackage{stackrel}
\usepackage{color}
\usepackage{comment}
\usepackage{bm}

\def \ee{\end{equation}}
\def \be{\begin{equation}}
\def \eea{\end{eqnarray}}
\def \bea{\begin{eqnarray}}

\begin{document}

\title{Padmanabhan's Boundary Variational Principle \\
for \\
Electrodynamics and Yang-Mills Theory
}
\author{Benjamin Koch}
\affiliation{Institut f\"ur Theoretische Physik,
 Technische Universit\"at Wien,
 Wiedner Hauptstrasse 8--10,
 A-1040 Vienna, Austria}
 \affiliation{ Instituto de F\'isica, Pontificia Universidad Cat\'olica de Chile, 
Casilla 306, Santiago, Chile}
\affiliation{Atominstitut, Technische Universit\"at Wien,  Stadionallee 2, A-1020 Vienna, Austria}
\date{\today}

\begin{abstract}
In this note, we revisit a variational principle introduced by Padmanabhan for describing gravitation using a field action composed solely of a boundary term. We demonstrate that this procedure can also be applied to derive Maxwell's and Yang-Mills equations. Additionally, we find that in this boundary approach, $\mathcal{CP}$-violating dual couplings and spontaneous symmetry breaking through gauge boson masses can emerge in a manner analogous to how the cosmological constant appears in the original gravitational context.
\end{abstract}


\maketitle

\tableofcontents

\section{Introduction}

Action principles are foundational in theoretical physics, serving as the starting point for deriving the equations of motion governing physical systems.
Applications of action principles extend beyond fundamental physics to areas such as statistical mechanics, where the action framework aids in understanding phase transitions and critical phenomena.

The common feature of these principles is that by specifying an action, a single scalar quantity whose extremization yields the dynamics of a system, one can systematically derive the laws of physics through the calculus of variations. This approach not only unifies various physical theories but also provides deep insights into the symmetries and conservation laws inherent in these systems, as articulated by Noether's theorem~\cite{Noether:1918zz}. 

The differences between the numerous versions of this principle lie in the 
choice of dynamical variables, actions, variations, and applications.
The most renowned version of this principle is the principle of least action, which is central to classical mechanics, and field theory. For instance, in classical mechanics, Hamilton's principle provides a pathway to Newton's laws, while in quantum mechanics, Feynman's path integral formulation relies on the sum over histories of action~\cite{Feynman:1948ur}. In field theory, the Lagrangian and Hamiltonian formulations underpin electromagnetism, General Relativity (GR), and the Standard Model of particle physics. 
In Maupertuis's version of the principle
the initial and final points are fixed, while the total time is not.
Schwingers quantum action principle~\cite{Schwinger:1951ex,Schwinger:1951xk}, contemplates variations in transition amplitudes.
Each of these formulations, offers unique advantages, such as facilitating quantization, simplifying the treatment of constraints or of gauge redundancies.

Almost a century after Hilbert's variational approach to GR~\cite{Hilbert:1915tx},
Padmanabhan extended the list of action principles by proposing a novel version
which he used to derive the equations of GR~\cite{Padmanabhan:2006cj}. This version is based on particularly chosen metric variations on the boundary of a four-dimensional space-time volume instead of generic variations in the volume itself. The approach provided a new perspective on the thermodynamical aspects of gravitational theories~\cite{Padmanabhan:2009vy,Padmanabhan:2007en}.
This research is to be seen in the broader context of holographic interpretations of gravity \cite{tHooft:1993dmi,Susskind:1994vu,Maldacena:1997re}.
In addition, it offered a different take on the cosmological constant problem~\cite{Weinberg:1988cp,Durrer:2007re,Padmanabhan:2007xy,Padmanabhan:2002ji,Sola:2013gha}.

\subsection{Research hypothesis and structure of the paper}

The research hypothesis of this note is: 

{\it{Padmanabhan's Variational Principle (PVP) is not just a peculiarity applicable to gravity.
Instead, it can be adopted such that it also allows one to derive field equations of other gauge fields, such as electrodynamics (ED) or Yang-Mills (YM) theory from this boundary perspective.}}
When exploring this hypothesis, I will not go into the thermodynamic aspects of the original variation proposed for gravity. Instead, I will restrict myself to a formal definition of the PVP and work with this definition.

In the following subsection \ref{subsec_PadVarPrinc} the PVP is reformulated in such a way that it is applicable to a broader class of theories.
Then, in 
\ref{subsec_EDYM}, the 
usual derivation of Maxwell's and Yang-Mills equations is summarized.
These theories are then revisited with the PVP approach section \ref{sec_PVPEDYM}.
The results are discussed in section \ref{sec_Disc}, with particular emphasis on the Cosmological Constant Problem (CCP), the strong ${\mathcal{CP}}$ problem, and spontaneous symmetry breaking. Conclusions are presented in section \ref{sec_Conc}.

\subsection{Padmanabhan's variational principle}
\label{subsec_PadVarPrinc}

In this subsection, we will briefly revisit Padmanabhan's variational principle for surface actions, introduced in~\cite{Padmanabhan:2006cj}. While this approach was originally developed in the context of gravity, we aim to present the main steps in a manner that can be straightforwardly generalized to other gauge theories.
\begin{itemize}
    \item[a)] Formulate a surface action, which can be written as a total derivative ${\mathcal{A}}_{kin}(g_{\mu \nu}, {\mathcal{V}})= \int_{\mathcal{V}} d^4 x \partial(\dots)$ and combine
    it with an interaction ${\mathcal{A}}_{int}$, which is not necessarily a pure surface term ${\mathcal{A}}_{tot}(g_{\mu \nu}, {\mathcal{V}})={\mathcal{A}}_{kin}+{\mathcal{A}}_{int}$ (see Eq. (6) in~\cite{Padmanabhan:2006cj}).
    To distinguish these surface actions from the usual bulk actions, we use the letter $\mathcal{A}$ instead of $\mathcal{S}$.
    \item[b)] Vary ${\mathcal{A}}_{tot}(g_{\mu \nu}, {\mathcal{V}})$ with respect to degrees of freedom which take the form of gauge transformations. For the case of gravity this is
    \be\label{eq_PadVar}
    \delta_P g_{\mu \nu}= \nabla_\mu \xi_\nu +\nabla_\nu \xi_\mu ,
    \ee
    with non-vanishing values on the surface $\partial {\mathcal{V}}$ (defined between (6) and (7) in~\cite{Padmanabhan:2006cj}).
    Here, it is important to note that (\ref{eq_PadVar}) with the field variation $\xi_\mu$ is actually NOT a gauge transformation (e.g. coordinate transformation $x^\mu \rightarrow x^\mu + \xi^\mu$), since other degrees of freedom (e.g. matter fields $\psi$) are not varied.
    \item[c)] Isolate the ``not-gauge'' gauge parameters $\xi^\mu$ in $\delta_P {\mathcal{A}}_{kin}$ and use the generalized Stoke's theorem to write
    the variation as boundary integral
    \be\label{eq_deltaPA}
    \delta_P {\mathcal{A}}_{tot}(g_{\mu \nu}, {\partial\mathcal{V}})= \int_{\partial {\mathcal{V}}} d^3x 
    (\dots )_{\mu \nu} (\xi^\nu n^\mu)-
\int_{\partial {\mathcal{V}}} d^3x 
    (\dots )_{\alpha\mu \nu}  ( \delta_P g^{\mu \nu} n^\alpha )  
    ,
    \ee
    where $n^\mu$ is the normal to the surface $\partial {\mathcal{V}}$ contains the volume ${\mathcal{V}}$ (see Eq. (7) in~\cite{Padmanabhan:2006cj}).
    \item[d)]
    Choose a particular class of boundaries $\partial {\mathcal{V}}_P$, which represents the physics of interest and which further intersects with a space-time point given by the coordinate $x^\mu$.
    In the particular case of Padmanabahn's gravitational example, this implies that both the normal $n^\mu$ and the variation $\xi^\mu$ are chosen to be null vectors which are proportional to each other and that $\xi^\mu$ is the Killing vector that generates the boundary which is a local Rindler horizon H.
   This choice ensures for example ``that $\delta_p g^{\mu \nu}=0$ on the Killing horizon H making the second term in Eq. (\ref{eq_deltaPA}) vanish.''(See page 8 in~\cite{Padmanabhan:2006cj}).
    In our examples later on, we will not neither have to impose that $g^{\mu \nu}=0$, nor impose that the normal vectors $n^\mu$ are null a priori, even though such a constraint can still be imposed later on.
    Now, after choosing a suitable class of boundaries, impose that the variations (\ref{eq_PadVar}) do not change the value of the action for this class of boundaries $\partial {\mathcal{V}}_P$
    \be\label{eq_PadvarPrinciple}
   \left.\delta_P {\mathcal{A}}\right|_{
   \forall \partial{\mathcal{V}}_P \cap x^\mu}
   \equiv 0.
    \ee
    This
    results in the equations of motion, which are a contraction with the normal vector $n^\mu$. For the gravitational case, the contracted equations read
    \be\label{eq_eomPad1}
\left.    (R^\mu_\nu-4 \pi T^\mu_\nu)\xi^\mu \xi_\nu\right|_{\forall \{\xi^\mu,\; \xi^2=0\}}=0,
    \ee
    as it is shown in Eq.(8) of~\cite{Padmanabhan:2006cj}.
    The equations are valid on the boundary $\partial {\mathcal{V}}$, but ``one can do this around every event in spacetime locally and hence this result should hold everywhere''~\cite{Padmanabhan:2006cj}.
    These equations are the result of the PVP. They can be further manipulated, such that they take the form of the usual dynamical equations of the theory.
\end{itemize}
As part of these manipulations, the vector \(\xi^a\) can be integrated out of equation (\ref{eq_eomPad1}) using the Bianchi identity, yielding:
\begin{equation}\label{eq_eomPad2}
G^\mu_\nu = 8\pi T^\mu_\nu + \Lambda_0 \delta^\mu_\nu,
\end{equation}
where $G_{ab}$ is the Einstein tensor and \(\Lambda_0\) is understood as an integration constant (see Eq. (9) in ~\cite{Padmanabhan:2006cj}). Clearly, equation (\ref{eq_eomPad2}) can also be derived from a conventional variational principle acting on the bulk Einstein-Hilbert action with a cosmological constant \(\Lambda\) as an additional coupling constant of the theory. 
However, if this coupling constant is generated by quantum corrections of matter degrees of freedom, which are expected to take the form $T_{\mu \nu}\rightarrow T_{\mu \nu}+\Lambda g_{\mu \nu}$, it is challenging to explain why the observed value of $\Lambda$ is so extremely small. This conceptual problem is known as the Cosmological Constant Problem (CCP). By emphasizing the distinction between {\it integration constants} (like \(\Lambda_0\)) and {\it coupling constants} (like \(\Lambda\)) Padmanabhan states that while such a change of perspective ``does not {\it {solve}} the cosmological constant problem, it changes its nature completely because the theory is now invariant under $T_{\mu \nu}\rightarrow T_{\mu \nu}+\Lambda g_{\mu \nu}$'' transformations~\cite{Padmanabhan:2006cj}. This invariance can be seen from equation (\ref{eq_eomPad1}), since $\xi^a$ is chosen to be a null vector.
Thus, the CCP becomes less pressing if one thinks of gravity as a boundary theory subject to the PVP, rather than a bulk theory with conventional variations.

\subsection{
Maxwell's Electrodynamics and Yang Mills theory}
\label{subsec_EDYM}

In this subsection we will shortly recapitulate the derivation of Maxwell's and Yang-Mills' equations
from the usual variational principle. The readers familiar with this formalism won't miss anything if they jump right into section \ref{sec_PVPEDYM}.
Maxwell's equations can be elegantly derived from an action principle in the covariant formalism. The starting point is the action for the electromagnetic field, given by:
\begin{equation}\label{eq_MaxBulk}
{\mathcal{S}}^M = -\frac{1}{4} \int d^4x \, F_{\mu\nu} F^{\mu\nu},
\end{equation}
where 
\be\label{eq_Fmn}
F_{\mu\nu} = \partial_\mu A_\nu - \partial_\nu A_\mu 
\ee
is the electromagnetic field tensor, and \( A_\mu \) is the four-potential. 
To distinguish this bulk action from the surface actions in the PVP, we use the letter $\mathcal{S}$ instead of $\mathcal{A}$.
To derive the equations of motion, we apply the usual variational principle in the bulk. We vary the action with respect to the four-potential \( A_\mu'=A_\mu + \delta A_\mu \):
\begin{equation}\label{eq_deltaaM}
\delta {\mathcal{S}}^M = -\frac{1}{2} \int d^4x \, F^{\mu\nu} \delta F_{\mu\nu}.
\end{equation}
Using \( \delta F_{\mu\nu} = \partial_\mu \delta A_\nu - \partial_\nu \delta A_\mu \), and integrating by parts, we obtain:
\begin{equation}
\delta {\mathcal{S}}^M = \int d^4x \, (\partial_\mu F^{\mu\nu}) \delta A_\nu.
\end{equation}
For the action to be stationary, \( \delta {\mathcal{S}}^M = 0 \) for arbitrary \( \delta A_\nu \), leading to the Euler-Lagrange equation:
\begin{equation}
\partial_\mu F^{\mu\nu} = 0.
\end{equation}
This is one set of Maxwell's equations, representing the homogeneous equations in the absence of sources.

To include sources, we add the interaction term \( \int d^4x \, J^\mu A_\mu \) to the action, where \( J^\mu \) is the four-current. The total action becomes:
\begin{equation}
{\mathcal{S}}^M_{tot} = -\frac{1}{4} \int d^4x \, F_{\mu\nu} F^{\mu\nu} + \int d^4x \, J^\mu A_\mu.
\end{equation}
Applying the usual variational principle to this action, we obtain:
\begin{equation}\label{eq_Maxusual}
\partial_\mu F^{\mu\nu} = J^\nu,
\end{equation}
which are the inhomogeneous Maxwell's equations. 

Yang-Mills theory generalizes Maxwell's theory to
non-Abelian gauge groups.
The action for a Yang-Mills field $A_\mu^a$ is given by:
\begin{equation}
{\mathcal{S}}^{YM} = -\frac{1}{4} \int d^4x \, F^a_{\mu\nu} F^{a\mu\nu},
\end{equation}
where \( F^a_{\mu\nu} = \partial_\mu A^a_\nu - \partial_\nu A^a_\mu + g f^{abc} A^b_\mu A^c_\nu \) is the field strength tensor \( f^{abc} \) are the structure constants of the gauge group, and \( g \) is the coupling constant.
The structure constants satisfy the commutation relations:
\begin{equation}
[T^a, T^b] = i f^{abc} T^c,
\end{equation}
with \( T^a \) being the generators of the gauge group.
It is convenient to define the covariant derivative 
\begin{equation}
D_\mu^{ab} = \delta^{ab} \partial_\mu  + g f^{abc} A^b_\mu.
\end{equation}
Repeating the same steps as previously in equations  (\ref{eq_deltaaM}-\ref{eq_Maxusual})
we obtain, after somewhat more algebra
\begin{equation}\label{eq_YMusual}
D_\mu^{ab} F_b^{\mu\nu} = j^\nu_a.
\end{equation}
These equations describe the dynamics of the non-Abelian gauge fields, extending the principles of gauge invariance to more complex symmetries.

\section{PVP for Electrodynamics and Yang-Mills theory}
\label{sec_PVPEDYM}

When implementing the PVP for Electrodynamics and Yang-Mills theory, we have to keep in mind that several aspects of these theories are quite different from the theory of general relativity.
For example, while in the gravitational case, there exists a remarkable algebraic identity between the bulk and the surface Lagrangian (see Eq. (4) in~\cite{Padmanabhan:2006cj}), I am not aware of such an identity
Electrodynamics and Yang-Mills theory.
Therefore, I will focus on the similarities of the theories, in particular on the similarities in the context of the research hypothesis.

\subsection{PVP for electrodynamics}

We keep $A^\mu$ as the dynamical gauge field and the field strength defined in (\ref{eq_Fmn}).
Now, we follow the procedure $a)-d)$,
adopting the steps from a metric field $g_{\mu \nu}$ as the dynamical variable to the vector potential $A^\mu$:
\begin{itemize}
    \item[a)] 
    A natural choice for a surface term in ED is
\be\label{eq_AsurM}
{\mathcal{A}}_{kin}^M(A^\nu, {\mathcal{V}})= 
\int_{\mathcal{V}} d^4 x \; \partial_\mu (A_\nu F^{\mu \nu}).
\ee
Note that this action can be written as
\be
{\mathcal{A}}_{kin}^M(A^\nu, {\mathcal{V}})= 
\int_{\mathcal{V}} d^4 x \;  \left(\frac{1}{2}F_{\mu \nu} F^{\mu \nu}-A_\mu \partial_\nu F^{\mu \nu}\right).
\ee
This shows, that the boundary term (\ref{eq_AsurM}) can also be obtained by subtracting a $\sim A_\mu \partial_\nu F^{\mu \nu}$ term
from the bulk Maxwell action (\ref{eq_MaxBulk}).
The interaction with an external  electric current $j^\mu$ is given by
\be\label{eq_Aint}
{\mathcal{A}}_{int}^M(A^\nu, {\mathcal{V}})= 
-\int_{\mathcal{V}} d^4 xA^\mu j_\mu.
\ee
The total surface action is then 
\be\label{eq_Asurtot}
{\mathcal{A}}^M_{tot}(A^\nu, {\mathcal{V}})={\mathcal{A}}_{kin}^M(A^\nu, {\mathcal{V}})+{\mathcal{A}}_{int}^M(A^\nu, {\mathcal{V}}).
\ee
\item[b)]
The Padmanabhan variation of the vector field is taken to be
\be\label{eq_PadVarA}
\delta_P A^\mu= \partial^\mu \phi,
\ee
in analogy to the variation (\ref{eq_PadVar}).
Here, $\phi$ is the scalar not-gauge function available in ED, which is taking the role of the $\xi^\mu$ from the gravitational case. The variation (\ref{eq_PadVarA}) is not a gauge transformation, since it keeps the matter fields that give rise to the current $j^\mu$ fixed. Whereas a gauge transformation e.g. in QED would imply a simultaneous transformation of
the Dirac matter field $\psi\rightarrow e^{i\phi}\psi$.
\item[c)]
Now, we vary ${\mathcal{A}}_{tot}^M(A^\nu, {\mathcal{V}})$
with respect to (\ref{eq_PadVarA}), using the fact that for this variation $\delta F^{\mu \nu}=0$
\bea\label{eq_deltaPEM}
\delta_P {\mathcal{A}}_{tot}^M&=&\int_{{\mathcal{V}}} d^4x\; 
\left[
\partial_\mu ((\partial_\nu \phi)F^{\mu \nu})
+(\partial_\nu \phi)j^\nu
\right]\\ \nonumber
&=&\int_{{\mathcal{V}}} d^4x\; \left[
\partial_\mu \partial_\nu (\phi F^{\mu \nu})-
\partial_\mu(\phi \partial_\nu F^{\mu \nu})-\partial_\nu (\phi j^\nu)- \phi \partial_\nu j^\nu
\right]\\ \nonumber
&=&\int_{{\mathcal{V}}} d^4x\; \left[
-\partial_\mu(\phi \partial_\nu F^{\mu \nu})+\partial_\mu (\phi j^\mu)
\right]\\ \nonumber
\delta_P{\mathcal{A}}_{tot}^M(A^\nu, \partial{\mathcal{V}})&=&-\int_{\partial {\mathcal{V}}} d^3x\; n_\mu\left[\partial_\nu F^{\mu \nu}-j^\mu
\right]\cdot \phi.
\eea
Above, we used the chain-rule,
the conservation law for the current, the asymmetry of the tensor $F^{\mu \nu}$,
and finally the generalized Stoke's theorem.
The vector $n^\mu$ is the normal to the three-dimensional space-time surface.
\item[d)]
When imposing the variational principle (\ref{eq_PadvarPrinciple}) 
we need to specify for which class of surfaces ${\mathcal{V}}_P$ it shall be imposed.
For any point $x^\mu$
we choose a surface which intersect with the point.
For this choice, (\ref{eq_PadvarPrinciple}) implies
\be\label{eq_eomM1}
\left.
n_\mu\left[\partial_\nu F^{\mu \nu}-j^\mu
\right]
\right|_{\forall \{x^\mu\} \in \partial {\mathcal{V}}}
=0.
\ee
Since the function $\phi$ is completely arbitrary, this equation has to hold at any point of the boundary $\partial{\mathcal{V}}$.
This equation is completely analogous to the gravitational equation (\ref{eq_eomPad1}).
\end{itemize}

In the gravitational case, Padmanabhan further manipulated (\ref{eq_eomPad1}), such that it took the more conventional form (\ref{eq_eomPad2}). Now, we want perform similar manipulations for the electromagnetic case.
For this, we need to concentrate on the choice of boundaries.
\subsection{Which volume, which boundary?}

When formulating step $d)$ of the PVP, the nature of the boundary $\partial {\mathcal{V}}$, and thus the form of the normal vector $n^\mu$, is crucial for the physical interpretation of the setup.
As we will see, different volumes ${\mathcal{V}}$ with different boundaries $\partial {\mathcal{V}}$ can yield different results.
The geometrical quantities ${\mathcal{V}}$ and $\partial {\mathcal{V}}$ have to be defined in terms of the observer and/or the physical system of interest. For example, in the case of gravity, Padmanabhan defines light-like vectors $\xi^\mu$, since he considers classes of volumes bounded by a Rindler horizon~\cite{Padmanabhan:2007en,Chakraborty:2015aja}.
In this paper, I will distinguish between two types of static contour.

\begin{figure}[ht!]
\centering
\includegraphics[width=0.8 \linewidth]{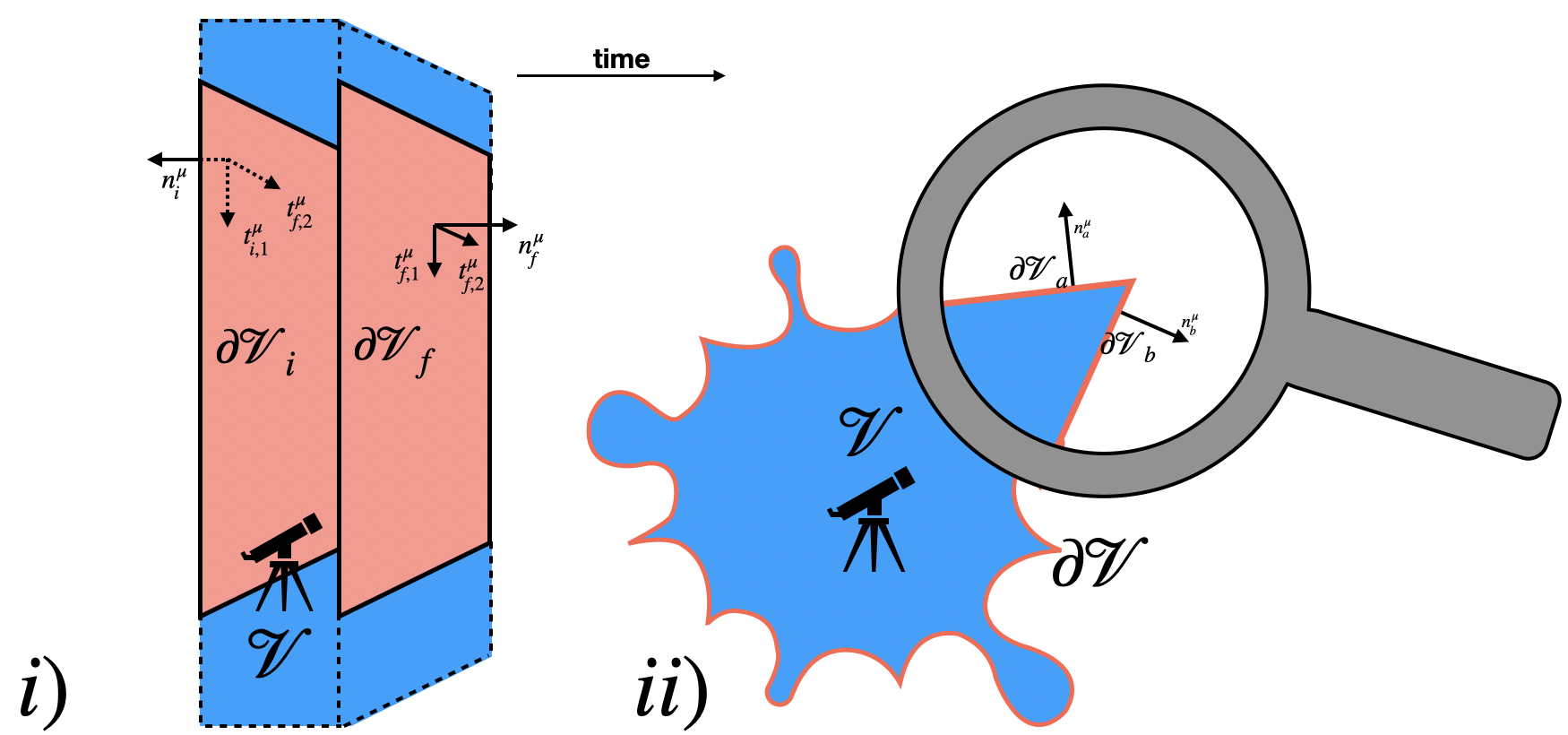}
\caption{
Static boundaries. Here, {\bf i)} corresponds to the {\bf{IF}} scenario with the time-like normal vectors $n_i^\mu$ and $n_f^\mu$ and the corresponding space-like boundaries $\partial{\mathcal{V}}_i$ and $\partial{\mathcal{V}}_f$.
The scenario {\bf{ASC}} is depicted in the right panel {\bf{ii)}}. The magnification glass exemplifies that such a scenario has to deal with numerous different normal vectors.
}
\label{splash}
\end{figure}

\begin{itemize}
    \item[{\bf IF)}] {\bf I}nitial and {\bf F}inal:\\
    As a first example, let's imagine an observer equipped with an infinite detector. He uses this detector to measure the field between two instants of time $t_i$ and $t_f$. In this case, the volume ${\mathcal{V}}$ is the spacetime volume between $t_i$ and $t_f$ in the rest frame of the observer. The boundary of this spacetime volume is given by the two three-dimensional patches $\partial{\mathcal{V}}_i$ at $t_i$ and $\partial{\mathcal{V}}_f$ at $t_f$. In a Minkowski diagram, this configuration looks like two infinitely large parallel plates, as indicated in figure~\ref{splash}-i). The time-like normal vectors to these surfaces are the anti-parallel four-vectors $n_i^\mu$ and $n_f^\mu$. We can further define three space-like tangential vectors $t^\mu_i$, which can be called $\hat x,\, \hat y, \, \hat z$. Equipped with these geometrical quantities, we can proceed to explore which equations give genuine solutions to the boundary equation of motion~(\ref{eq_eomM1}).

    The most general ansatz for such a solution that I propose is
\be\label{eq_eomM2b}
\partial_\nu F^{\mu \nu}-j^\mu+ B(A) n^\mu+
\epsilon^{\mu \nu \alpha \beta} D_{ \alpha\beta} n_\nu+E(A) A^\mu
+ F^i(A) t_i^\mu=0,
\ee
where the scalars $B$, $C$, $E$, and $F^i$ are functions of $A^\mu$, just like the antisymmetric tensor $D_{\nu \alpha}=D_{\nu \alpha}(A)$.
Contracting equation (\ref{eq_eomM2b}) with one of the normal vectors $n_\mu$, we get
\be\label{eq_eomM3}
n_\mu\left[\partial_\nu F^{\mu \nu}-j^\mu
\right]
+\left(
B n^2+ E (n \cdot A)\right)=0.
\ee
Let's specify $n^\mu$ as a time-like direction so that the result can be straightforwardly generalized to different configurations.
We note that there is no restriction on the scalars $F_i$. From the terms involving the normal vector, 
we can read off that 
\be\label{eq_B0}
B(A) = -\frac{E}{n^2} (n \cdot A).
\ee
Note that this condition implies that $B(A)$
cannot be a constant.
In the weak field limit, to leading order in $A^\mu$, the leading Lorentz-invariant representations for (\ref{eq_eomM2b}) and (\ref{eq_B0}) are
\bea\label{eq_BCDEleading}
D_{\mu \nu}&=&d^1 t_\mu^2 t_\nu^3+ d^2 t_\mu^3 t_\nu^1+ d^3 t_\mu^1 t_\nu^2+d^0 (\partial_\mu A_\nu-\partial_\nu A_\mu) +  {\mathcal{O}}(A^2)\\ \nonumber
E&=& E_0 +{\mathcal{O}}(A)\\ \nonumber
B &=& -\frac{E_0}{n^2} (n \cdot A)+{\mathcal{O}}(A^2)\\ \nonumber
F^i&=& F^i_0 + m^i_\mu A^\mu + {\mathcal{O}}(A^2),
\eea
where $d^\mu,\; m_\mu^i, E_0$, and $F_0^i$ are constants we can freely choose, like ``integration'' constants.
Note that even if the notation of these integration constants suggests otherwise, they are not necessarily four-vectors.
With this, the equation of motion (\ref{eq_eomM2b}) can be written as
\be\label{eq_eomM4}
\partial_\nu F^{\mu \nu}+
d^0 n_\nu \tilde F^{\mu \nu} 
=M_{\;\;\nu}^\mu A^\nu +\tilde j^\mu,
\ee
where we have grouped all terms which are field-independent into the observer-dependent source term
\be
\tilde j^\mu= j^\mu +B_0 n^\mu-
\epsilon^{\mu \nu \alpha \beta} (d^1 t_\alpha^2 t_\beta^3+ d^2 t_\alpha^3 t_\beta^1+ d^3 t_\alpha^1 t_\beta^2) n_\nu
- F^i_0 t_i^\mu.
\ee
On the right-hand side, we further note the appearance of a mass term for the field $A^\mu$, where the corresponding mass matrix is given by
\be\label{eq_massmatrix}
M^\mu_{\; \alpha} =E_0\left(\frac{1}{n^2}n_\alpha  n^\mu - \delta^\mu_\alpha\right)
+ (t^\mu_i m_\alpha^i ).
\ee
The first parenthesis of this mass matrix projects the fields $A^\mu$ onto directions that are tangent to the surface $\partial {\mathcal{V}}$.
Furthermore, this term breaks the $U(1)$ gauge symmetry.
Now let's use the fact that we defined the normal vector $n^\mu$ as the time direction.
In this case, equation (\ref{eq_eomM4}) simplifies significantly. For $\mu=0$ we find
\be
\partial_\nu F^{0\nu}=j^0.
\ee
Thus, Gauss's law is recovered without modification.
For $\mu = l$ we get
\be
\partial_\nu F^{l \nu}\pm d^0 \tilde F^{l \nu}= j^l\pm d^l- F_0^l-E_0 A^l + m^l_\alpha A^\alpha. 
\ee
The sign of two of the terms that modify Ampère's law depends on whether we choose $t_i$ or $t_f$. It is natural to impose that the dynamics should not depend on this choice, and the same equations should hold at any time. Thus, we obtain
\be
\partial_\nu F^{l \nu}= j^l- F_0^l-E_0 A^l + m^l_\alpha A^\alpha. 
\ee
This means that in the {\bf{IF}} configuration of boundaries, Ampère's law, obtained from the PVP, allows for an external shift of the current $j^l$ by $F_0^l$ and for a Lorentz-violating mass term introduced by the integration constants $E_0$ and $m_\alpha^l$.
These external modifications are reminiscent of a particular type of the large class of Lorentz violations introduced by Kostelecky~\cite{Kostelecky:2009zp}.

A completely analogous discussion can be repeated
for two two-dimensional parallel detector plates
which are located at two positions $x_i$ and $x_f$ and which collect data at all times $t$.

  \item[{\bf ASC)}]  {\bf A}rbitrary {\bf S}hape, but {\bf C}losed:\\
 As a second example, let's imagine an observer with smaller financial resources measuring between some initial time $t_i$
 and some final time.
 Neither is the size of the detector infinite, nor do all parts of the detector start their measurement at the same time. As long as the detector components take data continuously, this observer will be able to explore an ideally simply connected spacetime volume ${\mathcal{V}}$. However, the boundary of this volume will take a more complicated shape, as indicated by figure~\ref{splash}-ii).
 For such a complex spacetime volume, each segment of the boundary will have different normal vectors $n_{a,b,c...}^\mu$, as indicated
 by the magnification glass in figure~\ref{splash}-ii). For all these normal vectors, the equation of motion (\ref{eq_eomM1}) will have to be fulfilled. Assuming that the detector could also have been located at any other initial spacetime point, we have to conclude that for all globally defined normal vectors $n_{a,b,c...}^\mu$, the same local equation has to hold
 \be\label{eq_eomM1b}
n^{a,b,c,\dots}_\mu\left[\partial_\nu F^{\mu \nu}-j^\mu
\right]
=0.
\ee
This condition can only be fulfilled
if the expression in the parenthesis vanishes
 \be\label{eq_eomM1c}
\partial_\nu F^{\mu \nu}
=j^\mu.
\ee
Thus, in the {\bf{ASC}} scenario we recover 
the ordinary Maxwell equations.
\end{itemize}

\subsection{Yang-Mills theory}
\label{subsec_PVPYM}

Now, we follow the procedure $a)-d)$,
adopting the steps from a metric field $g_{\mu \nu}$ as the dynamical variable to the YM vector potential $A^\mu_a$, where $a$ labels the color degrees of freedom of this field. For this, we rely on the findings in the previous subsection, generalizing it to non-Abelian $SU(N)$ groups:
\begin{itemize}
    \item[a)] 
    First, the surface action (\ref{eq_AsurM}) is equipped with a self-interaction part yielding
\be\label{eq_AsurYM}
{\mathcal{A}}^{YM}(A^\nu, {\mathcal{V}})= 
\int_{\mathcal{V}} d^4 x \; \left[\partial_\mu (A_\nu^a F_a^{\mu \nu})-\frac{g}{2}f^{abc} A_\mu^b A_\nu^c F_a^{\mu \nu}-A_a^\mu j_\mu^a
\right].
\ee
Here, we have already incorporated the coupling to a color charge $j^\mu_a$, which is covariantly conserved
\be
D_\mu j_a^\mu\equiv \partial_\mu j_a^\mu + g f^{abc} A_\mu^c j^\mu_b=0.
\ee
\item[b)] 
The Padmanabhan variation of the YM vector field is taken to be
\be\label{eq_PadVarYM}
\delta_P A_\mu^a= \partial_\mu \phi^a+ g f^{abc} A_\mu^b \phi^c,
\ee
in analogy to the variations (\ref{eq_PadVar} and \ref{eq_PadVarA}).
Here, $\phi^a$ are the scalar not-gauge functions. This implies that the variation of the field strength is
\be
\delta_P F^a_{\mu \nu}= g f^{abc} F^b_{\mu \nu} \phi^c.
\ee
\item[c)]
Now, we vary (\ref{eq_AsurYM}) with respect to (\ref{eq_PadVarYM}) and isolate
the boundary contributions 
\be
\delta_P{\mathcal{A}}^{YM}= \int_{\partial \mathcal{V}} d^3 x \; 
n_\mu \left[ 
\partial_\nu F_a^{\nu\mu}+ g f^{abc}A_\nu^b F_c^{\nu \mu}-j_a^\mu
\right] + \int_{\mathcal{V}} d^4x (\dots) \phi^a.
\ee
\item[d)]
Next, we impose the variational principle by choosing functions \(\phi^a\) that vanish in the bulk of \(\mathcal{V}\), but are finite on the boundary. This yields the YM equations of motion, contracted with the normal vector
\be\label{eq_YMeom1}
n_\mu \left[ 
\partial_\nu F_a^{\nu\mu}+ g f^{abc}A_\nu^b F_c^{\nu \mu}-j_a^\mu
\right]=0.
\ee
\end{itemize}
Like discussed in the previous section,
these equations can be manipulated for different boundary choices.
Two examples are as follows.
\begin{itemize}
    \item For the {\bf{IF}} scenario in the weak field limit for non-null vectors $n^\mu$  this yields
\be\label{eq_eomYMgen}
D_\mu^{ab} F_b^{\mu\nu}+
d^0 n_\nu \tilde F^{\mu \nu}_a 
=M_{\;\;\nu}^\mu A^\nu_a +\tilde j^\mu_a,
\ee
where the mass matrix on the right hand side is defined in (\ref{eq_massmatrix}). Thus, these equations can take the form of Proca equations, together with some other Lorentz and gauge-symmetry breaking integration constants.
    \item 
    For the {\bf{ASC}} scenario, 
with many different normal vectors,
we recover the YM equations (\ref{eq_YMeom1}).
\end{itemize}

\section{Discussion}
\label{sec_Disc}

Let us address some aspects of the above findings in more detail.
\subsection{Gauge co-variance}

Clearly, there is an intimate relation between Padmanabhan's variations $\delta_P$
and the corresponding infinitesimal gauge transformations $\delta_G$.
By condition $b)$, their action on the gauge fields is identical, as summarized below for metric-, Yang-Mills-, and Photon-fields
\be\label{eq_deltaPG}
\delta_P \left\{g_{\mu \nu}, \;A^a_\mu,\; A_\mu\right\}\equiv
\left\{\nabla_\mu \xi_\nu+\nabla_\nu \xi_\mu,\; \partial_\mu \phi^a + g f^{abc} A_{\mu}^b \phi^c, \;\partial_\mu \phi\right\}=
\delta_G \left\{g_{\mu \nu}, \;A^a_\mu,\; A_\mu\right\}.
\ee
Padmanabhan emphasizes that, the difference lies in the condition
that his variations only acts on the gauge fields themselves, while gauge transformations also act on matter fields $\psi(x)$.
For simplicity let us take this field to be a scalar field. Thus, while $\delta_P \psi(x)=0$,
the corresponding gauge transformations
are
\be\label{eq_deltaG}
 \left\{
 \delta_G^{\xi^\nu},\;
  \delta_G^{\phi^a}
  ,\;\delta_G^{\phi}
 \right\}\psi(x)
 =
\left\{
\psi(x^\mu + \xi^\mu), \; 
e^{i \phi^a t^a}\psi, \; 
e^{i \phi} \psi
\right\},
\ee
where $t^a$ are the generators of the YM gauge group. We note that in the Abelian case, the $U(1)$ current $j^\mu$ is not modified under this transformation, while the transformation of kinetic terms of matter degrees of freedom such as $\bar \psi \gamma^\mu \partial_\mu \psi$ behave differently under $\delta_G$ and $\delta_P$. 
In particular, in the case of electromagnetic theory it would be interesting to explore in a next step the possible relations between the PVP and the edge modes/ entanglement entropy mentioned in~\cite{Donnelly:2014fua}.
In contrast to this, in the YM case and the gravitational case
the sources $j^\mu_a$ and $T^{\mu \nu}$
do transform non-trivially under (\ref{eq_deltaG}).
We note further that boundary actions
are typically not gauge invariant without further conditions. This also true in the context of holography~\cite{tHooft:1993dmi,Susskind:1994vu,Maldacena:1997re}, where the problem is fixed in terms of fall-off conditions on the field space and corresponding conditions on the allowed gauge transformations, resulting in the so called ``small gauge transformations''.
In the case of Padmanabhan's formalism applied to gravity, gauge invariance can be recovered by restricting to linear gauge transformations~\cite{Padmanabhan:2006cj}. In the electromagnetic case studied here, imposing gauge invariance is synonymous to imposing the equation of motion on the boundary~(\ref{eq_deltaPEM}).
These resulting equations of motion ((\ref{eq_eomM1}) for the electromagnetic case, or (\ref{eq_YMeom1}) for the YM case) are then explicitly covariant under the respective gauge transformations. The equations (\ref{eq_eomM4}, \ref{eq_eomYMgen}), which are subsequently derived from these relations contain arbitrary constants, which like usual integration constants, can break symmetries of a system of equations.

\subsection{Some famous problems}

The nature of this subsection is less mathematical and more speculative than those in the previous sections of the paper. Readers who prefer to avoid the jargon and handwavyness, which ``naturally'' arises in the context of these naturality problems, should skip this subsection and jump directly to the conclusions.

Applying the PVP to electromagnetism and YM theory with observers with {\bf{IF}} boundaries gives
rise to terms that, like the cosmological constant, are inflicted with 
conceptual problems on their own right. Therefore, we briefly discuss these three prominent problems: the cosmological constant problem, the strong ${\mathcal{CP}}$ problem, and the problem of massive gauge bosons. We then compare and contrast these problems, highlighting their similarities and differences.

\subsubsection{The cosmological constant problem}

The cosmological constant problem refers to the discrepancy between the observed value of the cosmological constant, \(\Lambda\), and theoretical predictions. If the cosmological constant is associated with the energy density of the vacuum, then its value should be determined by the quantum fluctuations of the fields in the universe. However, theoretical estimates based on quantum field theory suggest a value that is many orders of magnitude larger than what is observed. 
This vast difference poses a significant challenge to our understanding of vacuum energy and gravitation, and it is one of the most profound problems in modern theoretical physics~\cite{Weinberg:1988cp,Durrer:2007re,Padmanabhan:2007xy,Sola:2013gha}.
As mentioned previously, in~\cite{Padmanabhan:2007xy} Padmanabhan discussed this problem by proposing his PVP for gravity.
He showed within this approach the cosmological constant $\Lambda_0$, even though absent in his original equations, appears in the solutions as integration constant.
Then he argued that it is much less conflictive to fine-tune an integration constant than an effective coupling.

\subsubsection{The strong ${\mathcal{CP}}$ problem}

The strong ${\mathcal{CP}}$ problem arises in the context of QCD, the theory of the strong interaction.  While the electroweak theory allows for ${\mathcal{CP}}$ violation, QCD appears to be highly ${\mathcal{CP}}$-conserving, despite the fact that a ${\mathcal{CP}}$-violating term is allowed by the theory's symmetry structure.
Typically Yang Mills equations, such as those of QCD are
obtained from a volume action 
\be\label{eq_Svol}
S_{vol}= \int_{\mathcal{V}} d^4x \;
\left[
\frac{1}{4} F_{\mu \nu}^a F^{\mu \nu}_a+
\frac{\theta}{4} F_{\mu \nu}^a \tilde F^{\mu \nu}_a+ A_{\mu}^aj^\mu_a
\right],
\ee 
using a conventional variation $\delta A^\mu_a = \delta A^\mu_a$ and neglecting boundary terms.
In (\ref{eq_Svol}), the ${\mathcal{CP}}$ violation would be controlled by the so-called $\theta$ term.
This ${\mathcal{CP}}$-violating term, characterized by the angle \(\theta\), should contribute to the electric dipole moment of the neutron. However, experimental limits on the neutron's electric dipole moment suggest that \(\theta\) is extremely small~\cite{Abel:2020pzs}, 
posing a fine-tuning problem known as the strong ${\mathcal{CP}}$ problem.
Famous suggested solutions of this problem involve additional particles, called axions~\cite{Peccei:1977hh,Dine:1981rt}.

\subsubsection{Massive gauge bosons and 
the Higgs hierarchy problem}

The Higgs hierarchy problem, also known as the hierarchy problem, is a follow-up problem of the discovery of the massive W and Z gauge bosons~\cite{UA1:1983mne}. Since mass terms break the gauge invariance of YM theories, 
this observation posed a serious tension between experimental fact and a theoretical symmetry requirement. The reconciliation came in terms of a mechanism coined ``spontaneous symmetry breaking''~\cite{Higgs:1964pj,Englert:1964et}.
This model predicted another massive particle
which was discovered later on
\cite{ATLAS:2012yve,CMS:2012qbp}.
The Higgs hierarchy problem
concerns the large discrepancy between the mass of this particle (around $125$ GeV) and the Planck scale (around \(10^{19}\) GeV)~\cite{Arkani-Hamed:1998jmv}. Quantum corrections to the Higgs boson mass are proportional to the largest energy scales in the theory, which means that without fine-tuning, the Higgs mass should naturally be much closer to the Planck scale than the electroweak scale. This fine-tuning issue challenges our understanding of why the Higgs boson mass is so much lighter than the Planck scale and suggests that new physics may be required to stabilize the Higgs mass.

\subsubsection{Comparison of problems}

While the cosmological constant problem, the strong ${\mathcal{CP}}$ problem, and the Higgs hierarchy problem all involve fine-tuning and large discrepancies between scales, they arise in different contexts and have distinct characteristics:

\begin{itemize}
    \item The cosmological constant problem deals with the energy density of the vacuum and its gravitational effects, highlighting a discrepancy between theoretical predictions and cosmological observations.
     \item The strong ${\mathcal{CP}}$ problem is a fine-tuning issue within QCD, where the absence of observed ${\mathcal{CP}}$ violation suggests an unexpectedly small value for the \(\theta\) parameter. In contrast to the other two hierarchy problems, the value of $\theta$ is compatible with zero, while $\Lambda$ and $m_H$ are ``just'' much smaller than expected on theoretical grounds.
    \item The problem of massive gauge bosons consists in finding a way to reconciliate the existence of such mass terms with the concept of gauge symmetry. 
    It is intimately related to the Higgs hierarchy problem which is concerned with the stability of the Higgs boson mass $m_H$ against quantum corrections, pointing to a potential need for new physics to explain the observed electroweak scale.
\end{itemize}
Despite these differences, similarities and overlap between these problems have been studied. For example, an interplay between the CCP and the strong ${\mathcal{CP}}$ problem has been investigated in \cite{Weiss:1987ns,DiLuzio:2020wdo}. A relation between the CCP and the hierarchy problem has been studied in~\cite{Bezrukov:2007ep}
by making the Higgs field the cosmological inflaton. The interplay between the Higgs mass and the strong ${\mathcal{CP}}$ problem has been discussed in~\cite{Hall:2018let}.

\subsection{Three sides of the same coin?}

In the original version of the PVP, the gravitational equation of motion (\ref{eq_eomPad1}) is a contraction with normal (null) vectors of the boundary $\xi^\mu$ and it does not contain any cosmological term.
The cosmological term $\sim \Lambda_0$ only appears after some manipulations in (\ref{eq_eomPad2}), when it is shown that any solution of the latter equations is also a solution of the former.

In this paper, when we explore whether the PVP can also be applied to ED and YM theory instead of gravity we find an analogous situation. The equation equations of motion are
(\ref{eq_eomM1}) in the ED case and (\ref{eq_YMeom1}) in the YM case. They take the form of a vector equation contracted with a normal vector $n^\mu$ defined by the boundary $\partial {\mathcal{V}}$ and the only coupling is the minimal coupling to an external source $j^\mu$ (or $j^\mu_a$ respectively). We then show that these equations are solved by uncontracted vector equations depending on the chosen class of boundaries $\partial \mathcal{V}$. For {\bf {IF}} boundaries the vector equations are (\ref{eq_eomM2b}) and (\ref{eq_eomYMgen}) respectively. Interestingly, we can show in the weak field limit that each of these equations contains numerous constants. We discuss the role of these constants with a particularly interesting example, which we label $d^0$ and $E_0$. The constant $d^0$
implies a violation of ${\mathcal{CP}}$ symmetry and in the context of QCD it's value is intimately bound to the strong ${\mathcal{CP}}$ problem.
The constant $E_0$ induces a mass matrix for the 
vector fields $A_\mu$ and $A_\mu^a$. 
This mass breaks the gauge symmetry of the theory ``spontaneously'' when going from 
(\ref{eq_eomM1}) to (\ref{eq_eomM2b}) in the ED case, or from 
(\ref{eq_YMeom1}) to (\ref{eq_eomYMgen}) in the YM case.
In the Standard Model, masses of gauge bosons are, as explained above, introduced by another spontaneous symmetry breaking, which for the case of the Higgs mechanism leads to the Higgs hierarchy problem. For {\bf {ASC}} boundaries, no such breaking is allowed and the usual Maxwell and YM equations are recovered.

We can summarize that the PVP applied to boundary formulations of GR, ED, and YM theory provides the ``integration'' constants (e.g. $\Lambda_0$, $d_4, E_0, \dots$). In conventional volume formulations of
these gauge theories, these constants are interpreted as fundamental couplings of the theory and each of these couplings is intimately related to deep conceptual problems known as the
CCP, the strong ${\mathcal{CP}}$ problem, and the Higgs hierarchy problem.
In contrast, coming from the PVP, these constants form part of the space specifying the solution of the equations subject to particular classes of boundaries. In this sense, are just ``integration'' constants.

Thus, from the perspective 
of the PVP, the cosmological constant problem,
the strong ${\mathcal{CP}}$ problem and the problem of massive gauge bosons appear to be different sides of the same coin, since they all arise spontaneously as ``integration'' constants when solving the boundary equations of motion.

\section{Conclusion}
\label{sec_Conc}

In this paper, we investigate the hypothesis that: 
{\it{Padmanabhan's Variational Principle  is not just a peculiarity applicable to gravity.
Instead, it can be adopted such that it also allows to derive field equations of other gauge fields, such as Electrodynamics or Yang-Mills theory from this boundary perspective.}}  
For this purpose, we formulate 4 steps, $a)-d)$ that allow to adopt the PVP to other gauge theories. By explicitly deriving the equations of motion for ED (\ref{eq_eomM1}) and YM (\ref{eq_YMeom1})
we can confirm the hypothesis.

For these equations we contemplate
different classes of boundaries ({\bf IF, ASC}). For the former class of boundaries,
 we further obtain modified field equations for ED (\ref{eq_eomM4}) and YM (\ref{eq_eomYMgen}). 
These equations contain ``integration'' constants, which suggest that from the perspective of the PVP, the cosmological constant problem, the strong ${\mathcal{CP}}$ problem and the problem of massive gauge bosons
are of the same nature~\footnote{Note that we should not claim that this finding
allows to draw any conclusion concerning
the ``well-posedness''~\cite{Martin:2012bt,Hossenfelder:2018ikr} or the solution of these fundamental problems, but we can be confident that the PVP can offer a novel perspective on these issues.
}.
For the latter class of boundaries,
we recover the conventional Maxwell and YM equations.

In summary, given that gravity, electromagnetism, and Yang-Mills theory are different in many ways, it is remarkable and encouraging that the PVP still yields consistent and interesting results for all of these theories.

\section*{Acknowledgements}
Many thanks to H. Padmanabhan for consenting to coin the PVP after her father Paddy.
Further thanks to A.~Aggarwal, F. Ecker, K. K\"ading, M. Pitschmann, A. Riahina, R. Sedmik,  and H. Skarke for discussion and comments.

\section*{Appendix}

\subsection{Discrete symmetries}

Below, I summarize the transformation behavior of some fields and quantities under the discrete symmetries
\begin{table}[h]
    \centering
    \begin{tabular}{c|c|c|c|c|}
      &$n_0$, $\partial_0$  &$\vec n$,$\vec t$, $\vec \nabla$& $\vec E$ & $\vec B$  \\ \hline
     $\mathcal{P}$ &+&$-$&$-$&+ \\
    $\mathcal{C}$ &+&+&$-$&$-$ \\
     $\mathcal{CP}$ &+&$-$&$+$&$-$ \\
      $\mathcal{T}$ &$-$&+&$+$&$-$ \\
       $\mathcal{CPT}$ &$-$&$-$&$+$&$+$ \\
    \end{tabular}
    \caption{Transformation of basic fields and quantities, where $n^\mu=(n^0,\vec n)$.}
    \label{tab:CPbasic}
\end{table}
These transformations imply for the terms appearing in (\ref{eq_eomM4} and \ref{eq_eomYMgen}) e.g. in the spatial part
\begin{table}[h]
    \centering
    \begin{tabular}{c|c|c|c|c|c|}
      & $\partial_0 \vec E$  &$\vec \nabla \times \vec B$& $\theta \partial_0 \vec B$ & $\theta \vec \nabla \times \vec B_i$& $\vec j$  \\ 
      \hline
     $\mathcal{P}$  &$-$&$-$&$+$&$+$&$-$ \\
    $\mathcal{C}$ &$-$&$-$&$-$&$-$&$-$ \\
     $\bm{\mathcal{CP}}$ &$\bm{+}$&$\bm{+}$&$\bm{-}$&$\bm{-}$&$\bm{+}$\\
      $\mathcal{T}$  &$-$&$-$&$+$&$+$&$-$\\
       $\mathcal{CPT}$ &$-$&$-$&$-$&$-$&$-$ \\
    \end{tabular}
    \caption{Transformation properties under the discrete symmetries of terms that appear in the Maxwell and YM system.}
    \label{tab:CPMax}
\end{table}


\bibliography{PVP3}
\bibliographystyle{unsrt}

\end{document}